\documentclass[pra,aps,tightenlines,twocolumn,superscriptaddress,showpacs]{revtex4}

\usepackage{epsfig,amssymb,amsmath,bm}
\begin{document}

\title{Weak force detection with superposed coherent states.}

\author{W.~ J.~ Munro}\email{bill_munro@hp.com}
\affiliation{Hewlett Packard Laboratories,Bristol BS34 8HZ, UK}

\author{K.~ Nemoto}\email{nemoto@informatics.bangor.ac.uk}
\affiliation{School of Informatics, Dean Street, University of Wales,
Bangor LL57 1UT, UK}

\author{G.~J.~Milburn}
\affiliation{Institute for Quantum Information California Institute of
Technology, MC 107-81, Pasadena, CA 91125-8100,}
\affiliation{Centre for Quantum Computer Technology, University of
Queensland, Australia}

\author{S. L. Braunstein}
\affiliation{School of Informatics, Dean Street, University of Wales,
Bangor LL57 1UT, UK}

\date{\today}

\begin{abstract}
We investigate the utility of  non classical states of
simple harmonic oscillators, particularly a superposition of coherent
states, for sensitive force detection.  We find that like squeezed states
a superposition of coherent states allows displacement measurements at
the Heisenberg limit. Entangling many superpositions of coherent states offers a significant
advantage over a single mode superposition states with the same mean photon number.

\end{abstract}

\maketitle
\section{Introduction}
Non classical states of light have received considerable attention in the field
of quantum and atom optics. Many non-classical states of light  have been
experimentally produced and characterised. These
states include photon number states, squeezed states and certain entangled
states. There are a number of suggested, and actual, applications of these states
in quantum information processing including; quantum cryptography
\cite{Bennett:1992:1,Kempe:1999:1}, quantum
teleportation\cite{Bennett:1993:1,Bennett:1996:2,Braunstein:2000:2,%
Plenio:1998:3,Braunstein:2000:1,Braunstein:1998:2}, dense
coding\cite{Braunstein:2000:3} and quantum
communication\cite{Bennett:1999:3,Schumacher:1996:1,Schumacher:1996:2} to name
but a few. They have also been proposed for high precision measurements such as
improving the sensitivity of Ramsey fringe
interferometry\cite{Huelga} and the detection of
weak tidal forces due to gravitational radiation. In this
paper we consider how non classical states
of simple harmonic oscillators may be used to improve the detection
sensitivity of weak classical forces.

When a classical force, $F(t)$, acts  for a fixed
time on a simple harmonic oscillator,
with resonance frequency $\omega$ and mass $m$, it displaces
the complex amplitude of the oscillator in phase
space with the amplitude and phase of the
displacement determined by the time dependence of the force\cite{Braginsky92}.
In an interaction picture rotating at the oscillator frequency, the action
of the force is simply represented by the unitary displacement operator
\begin{equation}
D(\alpha)=\exp(\alpha a^\dagger -\alpha^* a)
\end{equation}
where $a,a^\dagger$ are the annihilation and
creation operators for the single mode
of the electromagnetic field satisfying $[a,a^\dagger]=1$,
and $\alpha$ is a complex amplitude which
determines the average field amplitude, $\langle a\rangle = \alpha$. For
simplicity we will assume that the force displaces the
oscillator in a phase space direction that is orthogonal to the
coherent amplitude of the initial state, which we take to
be real with no loss of generality. The displacement is
thus in the momentum quadrature, $\hat{Y}=-i(a-a^\dagger)$.
To detect the force we would need to measure this quadrature. If the
oscillator begins in a coherent state $|\alpha_0\rangle$,
($\alpha_0$ is real) the displacement
$D(i\epsilon)$ causes the coherent state to evolve to
$e^{i \epsilon \alpha_0}|\alpha_0+i\epsilon\rangle$. The signal is then
measured to be $S=\langle \hat Y_{out} \rangle= 2\epsilon$, while the
variance in the signal is given by
$V=\langle \hat Y_{out}^2 \rangle-\langle \hat Y_{out}
\rangle^2 =1$. The signal to noise ratio is hence
\begin{eqnarray}
 SNR=\frac{S}{\sqrt{V}}=2\epsilon
\end{eqnarray}
which must be greater than unity to be resolved (the measured signal must
be greater than the uncertainty of this quadrature in
a coherent state). Thus we find a
standard quantum limit for the weak force detection as
\begin{equation}
\epsilon_{SQL}\geq \frac{1}{2}\ .
\end{equation}

\section{Weak force detection with squeezed states.}

It is well known\cite{Caves} that this
limit may be overcome if the oscillator is first
prepared in a squeezed state (a uniquely quantum mechanical state)
for which the uncertainty in the momentum quadrature is reduced below the
coherent state level. For the case of an appropriately squeezed vacuum state
\begin{equation}\label{singlemodesq}
|\psi\rangle=\sqrt{1-|\lambda|^2} \sum_{n=0}^\infty \frac{\lambda^n
\sqrt{(2 n)!}}{n!}|2 n\rangle
\end{equation}
where the mean photon number is given by
\begin{equation}
\bar n=\lambda^2/(1-\lambda^2)
\end{equation}
and $\lambda={\rm tanh} r$ (with r being the squeezing parameter). A weak
force causes a displacement $D(i\epsilon/2)$ on the squeezed vacuum.
In this case the signal and variance for the measured momentum quadrature
is given by\cite{WallsMil94}
\begin{eqnarray}
S&=&\langle \hat Y_{out} \rangle= 2\epsilon
\\ V&=&\langle \hat Y_{out}^2 \rangle-\langle \hat Y_{out} \rangle^2=e^{-2 r}
\end{eqnarray}
and hence a signal to noise ratio of $SNR= 2\epsilon e^{r}$
The minimum detectable force is given by\cite{WallsMil94}
\begin{equation}
\epsilon\geq \frac{1}{2e^{r}}
\end{equation}
which for large squeezing corresponds to $\epsilon_{min}\geq 1/4\sqrt{\bar{n}}$.
We see that squeezing provides an increased sensitivity that scales as
$1/\sqrt{\bar{n}}$.

Following early work by Bollinger et al. \cite{Bollinger},
Huelga et al.\cite{Huelga} have shown that
quantum entangled states can be used to improve the sensitivity of
frequency estimation using Ramsey fringe
interferometry. Can entanglement be used to
improve the sensitivity for force detection?  To begin, let
us consider an entangled state of two harmonic oscillators, the two mode
squeezed state,
\begin{equation}\label{twomodesq}
|\psi\rangle=\sqrt{1-\lambda^2}\sum_{n=0}^\infty \lambda^n|n,n\rangle
\end{equation}
where $|n,n\rangle=|n\rangle_1\otimes|n\rangle_2$. The entanglement in this
state can be seen in a variety of ways. Most obviously it is an
eigenstate of the number difference operator $a_1^\dagger a_1-a_2^\dagger a_2$,
between the two modes, and in the limit of large squeezing, $\lambda\rightarrow
1$, a near eigenstate of phase sum\cite{MilBraun99}.
Alternatively we can consider the correlations between quadrature phase
operators. In the limit of large squeezing ($\lambda\rightarrow 1$), the state
approaches a simultaneous eigenstate of both $\hat{X}_1-\hat{X}_2$ and
$\hat{Y}_1+\hat{Y}_2$, which is the kind of state considered by Einstein
Podolsky and Rosen\cite{EPR}. This kind of correlation has been exploited by
Furasawa {\it et al.}\cite{Furasawa} to realise an experimental teleportation
protocol. With two oscillators, we need to specify how the weak force acts.
We will specify that the force acts independently on each oscillator.
To detect the force, consider a measurement of the joint physical quantity
described by the operator $\hat{Y}_1+\hat{Y}_2$. It is then straightforward to
show that the signal and variance of the measured result,
after the displacement, are given by
\begin{eqnarray}
S&=&\langle \hat Y_1+\hat Y_2 \rangle= 4 \epsilon
\\ V&=&\langle \left(\hat Y_1+\hat Y_2\right)^2 \rangle-\langle \hat
Y_1+\hat Y_2
\rangle^2=2 e^{-2 r}
\end{eqnarray}
which gives a a signal to noise ratio of $SNR= 2\sqrt{2}\epsilon e^{r}$. The
minimum detectable force is then $\epsilon\geq 1 /(2\sqrt{2}e^{r})$ which
is a $\sqrt 2$ improvement over the single mode squeezed state. For large
squeezing the minimum detectable force can be expressed in terms of the
total mean photon number for both modes. In this
limit $\epsilon_{min} \approx 1/(4\sqrt{\bar{n}_{tot}})$.
This is the same scaling as we found for a single mode squeezed state.
The apparent improvement due to entanglement is simply a
reflection of the fact that we have a two mode resource with double the
mean photon number.

For the two mode squeezed state with the measurement scheme chosen,
there is simple way to understand this result. The entangled two
mode squeezed state,(\ref{twomodesq}), is easily disentangled
by the application of a unitary operator of the
form $U=\exp(-i\pi(a_1^\dagger a_2+a_1 a_2^\dagger)/4)$,
which does not change the total energy. We will refer to this unitary
transformation as the beam splitter transformation as in the case
that the two oscillator modes correspond to optical fields modes,
this transformation describes the scattering matrix of an optical
beam splitter. The resulting state becomes a (disentangled) product
state of two single mode squeezed states (as in Eq.(\ref{singlemodesq})).
The weak force now acts to displace each of the single mode squeezed states,
each of which may be used to achieve the squeezed state limit for displacement
detection. As there are two realisations of the measurement scheme
there will be an additional $1/\sqrt{2}$ improvement in sensitivity simply from
classical statistics. It is thus inaccurate to attribute the
improved force sensitivity of a two mode squeezed state to entanglement when
$\hat{Y}_1+\hat{Y}_2$ measurements are performed. In assessing the
limits to force detection using entangled states of N harmonic
oscillators we thus need to consider if any apparent improvement
could have been achieved simply by using  N copies of an appropriate
non classical state of a single harmonic oscillator.

Of course it may not always be so obvious to transform an entangled state
to a product of non classical states. Consider an entangled state of the form
\begin{eqnarray}\label{correlatedpair}
| \Psi \rangle = \sum_{n=0}^{\infty}c_n | n , n \rangle
\end{eqnarray}
This state is correlated in number, but unlike the two mode squeezed
state, it is not necessarily a near eigenstate of phase sum. If we
consider a measurement of $Y_1+Y_2$ as previously,  the signal and
variance after the displacement are
\begin{eqnarray}
S&=& 4 \epsilon\\
V&=&2 \left(1+\langle a^\dagger a+b^\dagger b\rangle-
\langle a^\dagger b^\dagger+a b\rangle \right)
\end{eqnarray}
which gives an improvement in the signal to noise ratio when
$ \langle a^\dagger a+b^\dagger b\rangle < \langle a^\dagger b^\dagger +a
b\rangle$. A state like this, with correlated photon number,
is the pair-coherent (or ``circle'') state given by\cite{A86-827,RK93-552}
\begin{eqnarray}
|\mbox{circle}\rangle_{m} = {\mathcal{N}}
\int_{0}^{2\pi} |\alpha e^{i\varsigma}\rangle_{a}
  |\alpha e^{-i\varsigma}\rangle_{b} d\varsigma
  \label{eqn:circlestate}
\end{eqnarray}
where $|{\ldots}\rangle_{a}$ and $|{\ldots}\rangle_{b}$ represent coherent
states in the modes $\hat a$ and $\hat b$.  ${\mathcal{N}}$ is a
normalisation coefficient and $\alpha$ the amplitude of the coherent state.
This state can be written in the form (\ref{correlatedpair}) with
\begin{eqnarray}\label{circlecn}
c_{n}=   \frac{1}{\sqrt{I_{0}\left(2 \alpha \right)}} \frac{\alpha^{n}}{n!}.
\end{eqnarray}
Here $I_{0}$ is a zeroth order modified Bessel function. This state
cannot be separated into product states via beam splitter transformations.
It is easily shown that the minimum detectable force occurs when
\begin{eqnarray}
\epsilon_{min}=\frac{1}{2}\sqrt{\frac{1}{2}+\bar n - \alpha}
\end{eqnarray}
with the mean photon number being given by $\bar n=\alpha I_{1}\left(2 \alpha
\right)/I_{0}\left(2 \alpha \right)$. A small improvement is seen for all
$\alpha$, with the minimum occurring at $\alpha=0.85$
($\epsilon_{min}=0.221108$). As $\alpha\rightarrow \infty$ we have
$\epsilon_{min}\rightarrow 0.25$. In this optimal region the mean photon number
is small. The measurement of $Y_1+Y_2$ is not optimal however because
it is not a near eigenstate.

It is likely that one can achieve a significantly better sensitivity 
if one changes their measurement quantity from $Y_1+Y_2$ to a 
quantity that is a near eigenstate of (\ref{correlatedpair}). 
For these correlated photon number systems this could require a 
measurement of the photon number difference of (\ref{correlatedpair}) 
which with current technology is quite unpractical.

\section{Weak force detection with cat states.}

Let us now turn our attention to a less straightforward example.
In the previous example two entangled harmonic modes, the two mode
squeezed state, gave an improvement in the signal to noise ratio (compared to a
single mode) of $\frac{1}{\sqrt{2}}$.  With an entangled state comprised
of more modes, an even better improvement may be achievable. However there is no
simple way to generalise the two mode squeezed state to
give an entangled state of many modes. We now consider
another class of non classical states, based on a coherent superposition of
coherent states, which can be entangled over $N$ modes.


Consider $N$ harmonic oscillators prepared in the cat state
\begin{equation}
|\psi\rangle_N={\cal
N}_+(|\alpha,\alpha,\ldots,\alpha\rangle+|-\alpha,-\alpha,\ldots,-\alpha\rangle)
\label{ggcat}
\end{equation}
where
\begin{equation}
|\alpha,\alpha,\ldots,\alpha\rangle=\Pi^{\otimes N}_k|\alpha\rangle_k
\end{equation}
is tensor product of coherent states and ${\cal N}$ is the normalisation
constant given by
\begin{equation}
{\cal N}= \frac{1}{\sqrt{2+2 e^{-2 N |\alpha|^2 }}}
\end{equation}
We take $\alpha$ to be real for convenience. For $\alpha>>1$ this normalisation
constant approaches $1/\sqrt{2}$, and we henceforward make this assumption.
Parkins and Larsabal\cite{Parkins2000} recently suggested how this highly
entangled state might be formed in the context of cavity QED and quantised
motion of a trapped atom or ion.

To begin our consideration of these states, let us consider the case of a
single oscillator ($N=1$),
\begin{equation}
|\phi\rangle =\frac{1}{\sqrt{2}}\left
(|\alpha\rangle+|-\alpha\rangle\right )
\label{even_cat}
\end{equation}
where the mean photon number is given by $\bar n=|\alpha|^2$.
When a weak classical force acts on the state in Eq.(\ref{even_cat}) it is
displaced by
\begin{eqnarray}
|\phi\rangle_{out}&=&
\frac{1}{\sqrt{2}}\left(e^{-i Im
( \alpha \beta^\ast)}|\alpha+\beta\rangle+e^{i Im
( \alpha \beta^\ast)}|-\alpha+\beta\rangle\right ) \nonumber \\
&\approx& \frac{1}{\sqrt{2}}\left(e^{i \theta}|\alpha\rangle+e^{-i
\theta}|-\alpha\rangle\right ) \nonumber \\
& = &   \cos\theta|+\rangle+i\sin\theta|-\rangle
\end{eqnarray}
where $\theta=-Im ( \alpha \beta^\ast)$ and we have defined the even
($|+\rangle$) and odd parity ($|-\rangle$) eigenstates
\begin{equation}
|\pm\rangle=\frac{1}{\sqrt{2}}\left (|\alpha\rangle\pm|-\alpha\rangle\right )
\end{equation}
Our problem is thus reduced to finding the optimal readout for the rotation
parameter $\theta$ for a two dimensional sub-manifold of parity eigenstates.
The rotation is described by the unitary transformation
\begin{equation}
U(\theta)=\exp\left (i\theta\hat{\sigma}_x\right )
\end{equation}
where $\hat{\sigma}_x=|+\rangle\langle -|+|-\rangle\langle +|$ is a Pauli
matrix.

The objective is now to find an optimal measurement scheme to
estimate the rotation parameter, $\theta$, and thus the force
parameter, $\epsilon$. The maximum sensitivity will
occur when $\theta=-Im ( \alpha\beta^\ast)$ is
maximised for a given displacement.
Having chosen $\alpha$ real, $\theta$ is maximised by choosing
$\beta$ purely imaginary. This corresponds to a displacement $D(\beta)$
entirely in the momentum quadrature. Setting $\beta=i\epsilon$, we have
$\theta=\epsilon\alpha$. The theory of optimal parameter
estimation\cite{BCM} indicates that the limit on
the precision with which the rotation parameter can be determined is
\begin{equation}\label{parameter}
(\delta\theta)^2\geq\frac{1}{Var(\hat{\sigma}_x)_{in}}
\end{equation}
where $Var(\hat{\sigma}_x)_{in}$ is the variance in the generator of
the rotation in the input state $|+\rangle$, which is simply unity.
Thus we find that uncertainty on the force parameter is bounded below by
$\delta\epsilon\geq 1/(2\alpha)$. It thus follows that the minimum
detectable force is $\epsilon_{min}\geq 1/(2\alpha)$,
which may be written in terms of the total mean excitation number of the
input state as
\begin{equation}
\epsilon \geq \frac{1}{2\sqrt{\bar n}}
\end{equation}
where the mean photon number $\bar n=|\alpha|^2$.
This measurement is at the Heisenberg limit.
Comparison with the result for the single mode squeezed
state shows a similar dependence on the mean excitation number
however the squeezed state sensitivity is better
by a factor $1/2$.

We can now consider a two mode entangled cat state.
\begin{equation}
|\psi\rangle_1={\cal N}(|\alpha,\alpha\rangle+|-\alpha,-\alpha\rangle)
\end{equation}
However this state is easily disentangled with the unitary transformation
\begin{equation}
U(\pi/2)=\exp[-i\frac{\pi}{2}(a^\dagger_1 a_2+a_1 a^\dagger_2)]
\end{equation}
(for a quantum optical realisation this is a 50:50 beam-splitter) to produce
the separable state
\begin{eqnarray}
|\tilde{\psi}\rangle_1
& = & {\cal N}_1{\cal N}_2\left
(|\alpha\rangle_1+|-\alpha\rangle_1\right )\otimes\left
(|\alpha\rangle_2+|-\alpha\rangle_2\right )
\end{eqnarray}
As in the case for squeezed states, we only need consider the force
detection sensitivity for the state of a single oscillator.
The minimum detectable force is given by
\begin{equation}
\epsilon\geq \frac{1}{2\sqrt{2 \bar n}}
\end{equation}
Here we see the $\sqrt 2$ improvement from classical averaging. For the $N$ mode
state given by eqn (\ref{ggcat}) a linear transformation also exists to
transform the $N$ mode entangled state to a product state of single mode cat
states. In this case the minimum detectable force using $N$ modes, each
prepared in cat state with amplitude $\alpha$, is
\begin{equation}
\epsilon_{min}>\frac{1}{2\sqrt{N\bar n}}
\end{equation}
As each mode has a mean photon number given
by $\bar n=\alpha^2$, the total mean photon number used in the entire experiment is
$\bar{n}_{tot}=N\alpha^2$, the minimum detectable force can be written as
$\epsilon_{min}>1/\sqrt{\bar{n}_{tot}}$. We see from here that there is no real advantage in
using entangled states with the measurement protocol outlined, as the improvement is only the
standard statistical improvement that one gets from multiple copies of a single mode cat state
produced by disentangling the state.

\section{Entangled Cat states}

A question that be asked is whether both entanglement and collective
measurements allow one to increase the sensitivity of this displacement
measurement past the limits shown above? To address this question let us
consider again the  $N$ mode entangled cat state
\begin{equation}\label{nmodecat}
|\psi\rangle=\frac{1}{\sqrt{2}}(|\alpha,\alpha,\ldots, \alpha\rangle+
|-\alpha,-\alpha,\ldots,-\alpha \rangle)
\end{equation}
where the total photon number of the entire state is $n_{tot}=N \alpha^2$.
The weak force acts simultaneously on all modes of this $N$ party entangled
cat state. It causes a displacement $D(i \epsilon)$ on each mode in (\ref{nmodecat})
resulting in the state
\begin{eqnarray}\label{fulldisplaced}
|\psi (\theta)\rangle&=&\frac{e^{i N \theta}}{\sqrt{2}} |\alpha+i \epsilon,\alpha+i
 \epsilon,\ldots,\alpha+i \epsilon\rangle  \\
&+& \frac{e^{-i N \theta}}{\sqrt{2}}|-\alpha+i \epsilon,-\alpha+i \epsilon,\ldots,
-\alpha+i \epsilon \rangle \nonumber
\end{eqnarray}
where $\theta= \epsilon \alpha$. The theory of optimal parameter estimation indicates that the
limit on the precision with which the rotation parameter is given by (\ref{parameter}) but
where $\sigma_x=\sum_{i=1}^{N} \sigma_{x_i}$. The uncertainty in this force parameter is hence
bounded by
\begin{eqnarray}
\epsilon&=&\frac{1}{N \alpha} =\frac{1}{\sqrt{N n_{tot}}}
\label{entangled_emin}
\end{eqnarray}
and is at the Heisenberg Limit. We observe a critically important extra $\sqrt{N}$
improvement due to the entangled state and collective measurement (projective measurements
onto $|\alpha,\alpha,\ldots, \alpha\rangle-|-\alpha,-\alpha,\ldots,-\alpha \rangle$)
which can be seen over $N$ individual copies of the state $|\alpha\rangle+|-\alpha \rangle$, or
a single mode state $|\sqrt{n_{tot}}\rangle+|-\sqrt{n_{tot}}\rangle$. For a large and finite
$n_{tot}$ it seems optimal from that one should create highly entangled cat
state with as many modes as possible while maintaining $\alpha \gg 1$.

In our consideration so far we have not considered the effects of
loss or decoherence on these highly non-classical states. Whether we are considering highly
entangled cat states or large amplitude single mode cat states these are all extremely
sensitive to small amounts of loss and decoherence. Error correction and avoidance
techniques can be employed to reduce these effects but are beyond the scope of this paper.

\section{Generalised cat states.}
In the example just discussed, maximum sensitivity required the classical
force to displace the cat states in a direction orthogonal
to the phase of the superposed coherent amplitudes.
In general there is no way to arrange this before hand, as the phase of the
displacement depends on an unknown time dependence of the classical force.
However  a simple generalisation of the previous
cat states can be used to relax this constraint.
Note that the cat states are parity eigenstates and are thus the conditional
states resulting from a measurement of the number operator modulo 2,
$\hat{n}_N=a^\dagger a\  \mbox{mod}\ 2$, on an input state
$|\alpha\rangle$ with $\alpha$ real. We are thus led to consider the
conditional states for measurements of
$\hat{n}_K=a^\dagger a\  \mbox{mod}\ K$. Such states have previously been
considered by Schneider et al.\cite{Schneider98}.
Given a result $\nu=0,1,\ldots, K-1$ for such a measurement, the
conditional (unnormalised) states are
\begin{equation}
|K,\nu\rangle = \sum_{\mu=0}^{K-1}\exp\left [\frac{2\pi i\mu\nu}{K}\right
]|\alpha e^{2\pi i\mu/K}\rangle
\end{equation}
which are eigenstates of $e^{i2\pi a^\dagger a/K}$ with eigenvalues $e^{-i
2\pi\nu/K}$.

The case of $K=4$ has recently been considered by Zurek\cite{Zurek01} in
the context of decoherence and quantum chaos.
Assume that the oscillator is initially prepared in the state
\begin{equation}
|\psi\rangle_{in}=|4,0\rangle=|\alpha\rangle+|i\alpha\rangle+|-i\alpha\rangle+|-
\alpha\rangle
\end{equation}
with $\alpha$ real.
Under the action of a weak force characterised by a complex amplitude
displacement $\beta$, the output state is
\begin{equation}
|\psi\rangle_{out}=e^{i\theta}|\alpha\rangle+e^{i\phi}|i\alpha\rangle+e^{-i\phi}
|-i\alpha\rangle+e^{-i\theta}|-\alpha\rangle
\end{equation}
where $\theta=\alpha \mbox{Im}(\beta)$ and $\phi=\alpha\mbox{Re}(\beta)$.
The state now carries information on both
the real and imaginary components of the  displacement due to the force
which may be extracted by measuring the projection operator onto
the initial state. In the limit that $K\ >>\ |\alpha|^2\ >>1$,
the initial conditional state is simply the vacuum state and we
recover the usual standard quantum limit for force detection by number
measurement\cite{Caves}.

\section{Discussion and Conclusion.}
We now compare our results to the study of Ramsey fringe interferometry
introduced by Bollinger et al. \cite{Bollinger} and discussed by Huelga et
al.\cite{Huelga}. In Ramsey fringe interferometry the objective is to
detect the relative phase difference between two superposed states,
$\{|0\rangle,|1\rangle\}$. that form a basis for a two dimensional Hilbert
space. These states could be the ground and excited states of an electronic
dipole transition. The problem reduces to a quantum parameter estimation
problem. The unitary transformation which induces a relative phase in the
specified basis is $U(\theta)=\exp[i\theta \hat{Z}]$
where $\hat{Z}=|1\rangle\langle 1|-|0\rangle\langle 0|$. We are
free to choose the input state $|\psi\rangle_i$ and the
measurement we make on the output state, which is
described by an appropriate positive operator
valued measure (POVM).

The theory of quantum parameter estimation\cite{BCM} indicates in this case
that we should choose the input state as
$|\psi\rangle_i=(|0\rangle+|1\rangle)/\sqrt{2}$ and the optimal
measurement is a projective measurement in the basis
$|\pm\rangle=|0\rangle\pm|1\rangle$. The probability to obtain the result
$+$ is  $P(+|\theta)=\cos^2\theta$. In $N$ repetitions of the measurement the
uncertainty in the inferred parameter is
\begin{equation}
\delta\theta=\frac{1}{\sqrt{N}}
\end{equation}
which achieves the lower bound for quantum phase parameter estimation.
Repeating the measurement $N$ times is equivalent to a single product
POVM on the initial product state
$\prod_{i=1}^N\otimes(|0\rangle_i+|1\rangle_i)/\sqrt{2}$. However
it was first noted by Bollinger et al.\cite{Bollinger}  that
a more effective way to use the $N$ level systems is to
first prepare them in the maximally entangled state.
\begin{equation}
|\psi\rangle=\frac{1}{\sqrt{2}}(|0\rangle_1|0\rangle_2\ldots|0\rangle_N+|1\rangle
_1|1\rangle_2\ldots|1\rangle_N)
\label{entangle}
\end{equation}
and subjecting the entire state to the unitary transformation
$U(\theta)=\prod_{i=1}^N\exp(-i\theta\hat{Z}_i)$, the uncertainty in the
parameter estimation then achieves the Heisenberg lower bound of
\begin{equation}\label{limit}
\delta\theta=\frac{1}{N}
\end{equation}
Briefly let us instead consider $N/2$ maximally entangled pairs.
In this case we can combine (\ref{limit}) at $N=2$ with the square
root statistical benefit of $N/2$ repetitions  This yields
$\delta \theta = 1/2 \times \sqrt{2/N} = 1/\sqrt{2N}$
indicating that pairwise entanglement yields only a
margined benefit compared to full N-wise entanglement of for the phase
estimation.

We will now show that the entangled state in Eq.(\ref{entangle}) is in fact a
cat state for a collective operator algebra. The Hilbert space of $N$ two level
systems is the tensor product space of dimension $2^N$. The entangled state
in Eq.(\ref{entangle}) however resides in a
lower dimensional subspace of permutation symmetric
states\cite{Dicke}. These states constitute an $N+1$ dimensional irreducible
representation of SU(2) with infinitesimal generators defined by
\begin{eqnarray}
\hat{J}_z = \frac{1}{2}\sum_{i=1}^N\hat{Z}_i,\;\;
\hat{J}_y = \frac{1}{2}\sum_{i=1}^N\hat{Y}_i,\;\;
\hat{J}_x =\frac{1}{2}\sum_{i=1}^N\hat{X}_i
\end{eqnarray}
where $\hat{Z}_i=|1\rangle_i\langle 1|-|0\rangle_i\langle 0|,\
\hat{X}_i=|1\rangle_i\langle 0|+|0\rangle_i\langle 1|,
\hat{Z}_i=i|1\rangle_i\langle 0|-i|0\rangle_i\langle 1|$. The Casimir
invariant is $\hat{J}^2=\frac{1}{4}(\hat{J}_x^2+\hat{J}_y^2+\hat{J}_z^2)$
with eigenvalue $\frac{N}{2}(\frac{N}{2}+1)$. The operator $\hat{J}_z$ has
eigenvalues $m=-N/2, -N/2+1,\ldots,N/2$ which
is one half the  difference between the number of zeros and ones in an
eigenstate. It is more convenient to use
of the eigenstates, $|m\rangle_{N/2}$, of these commuting operators as basis
states in the permutation symmetric subspace. In this notation
the entangled state defined in Eq.(\ref{entangle}) may be written
\begin{equation}
|\psi\rangle=\frac{1}{\sqrt{2}}(|-N/2\rangle_{N/2}+|N/2\rangle_{N/2})
\end{equation}
In this form we can regard the state as an SU(2) `cat state' for $N$
two-level atoms. Hence it is straightforward to see that a single
$2^N$ level atom can achieve the same frequency sensitivity. 
Their equivalence can be also be understood by noting that 
the sensitivity of such frequency measurements is
proportional to the energy difference of the states involved. What
entanglement allows is for one to create an effective state 
without the need of resorting to create a superposition between 
certain ground state and a highly excited one.
 
A closer atomic analogy to a single mode cat state would be a cat state for
a single N level electronic system. For example we could consider the unnormalised
state defined on a hyperfine manifold with quantum number F, $|F\rangle_F+|-F\rangle_F$.
Such states have been considered in reference \cite{Xu99}. A similar state
could also be generated for the large magnetic molecular systems considered in
references \cite{Babara,Gider,Tejada}. The key is in how the resources can be
distributed and what type of measurements one is trying to achieve. If the
single molecule can only be prepared with a certain $N$, then an advantage
can be gained for frequency measurements by entangling the state of many
single molecule systems\cite{Tejada,Tejada1}. However if we restrict the total
system to having a fixed $N$ and we have enough control so as to be able to
prepare the system either as a single large SU(2) molecular state or
many entangled smaller molecular states, then the same sensitivity is
achieved for the high precision frequency measurement (this was not the
case for weak force measurements).

{\bf To conclude}, we have in this article shown how superpositions of
coherent states can be used to achieve extremely sensitive force
detection. For a single mode state $|\alpha\rangle + |-\alpha\rangle$ we have found that the
minimum detectable displacement for weak force measurements scales inversely proportional to
the square root of the mean photon number of the superposition of coherent states. This is
the same scaling obtained by a single mode squeezed state and
achieves the Heisenberg limit for single mode displacement measurements.

What is potentially more interesting is that if we take a number of individual copies of a
single mode cat states then we still achieve the inverse square root scaling
with total mean photon number (hence effectively allow one to increase the mean number of 
particles). If one starts with an $N$ mode entangled cat state, then
simple linear transformation can be used to turn this state into $N$ copies of a single mode
cat state and hence achieve the $\epsilon_{min}\sim {1}/{\sqrt{n_{tot}}}$ sensitivity.
This however is not optimal as it does not achieve the Heisenberg limit for multiple modes.
To achieve this limit for weak force detection one must use both an entangled $N$ mode cat 
state and a joint collective measurement (between the various modes). 
Entanglement is the critical resource to achieve the best sensitivity for a fixed $n_{tot}$. 
On the other hand we have shown for frequency (phase) measurements that the sensitivity previously
offered by entangling $N$ two levels atoms can be achieved with a single
$2^N$ level atom. The key is that the sensitivity for the frequency measurement is 
proportional to the energy difference of the states involved and both the entangled resource 
and the superposition resource have the same energy difference.
Entanglement allows one to create an effective state without the need of 
resorting to create a superposition between  certain the ground state 
and a highly excited one.

\section*{Acknowledgements}
GJM acknowledges the support of the Institute for Quantum Information,
California Institute of Technology where this work was initiated. WJM. KN and
SLB acknowledges funding in part by the European projects
EQUIP(IST-1999-11053), QUICOV and QUIPROCONE (IST-1999-29064).
The Australian Research Council Special Research Centre for Quantum Computer
Technology also supported this work.

\end{document}